\documentclass[a4paper,11pt,oneside,final]{article}

\usepackage[utf8]{inputenc}
\usepackage[T1]{fontenc}
\usepackage{amsmath}
\usepackage{amsfonts}
\usepackage{amssymb}
\usepackage{graphicx}
\usepackage{hyperref}
\usepackage[top=3cm, bottom=2.5cm, left=3cm, right=3cm]{geometry}

\title{Photonic jet: direct micro--peak machining}

\author{Robin Pierron\footnotemark[2] \footnotemark[1] , Pierre Pfeiffer\footnotemark[2] , Grégoire Chabrol\footnotemark[2] \footnotemark[3] , Sylvain Lecler\footnotemark[2]}

\date{Article published in \textit{Applied Physics A} 123, 686 (2017) \\ 
https://doi.org/10.1007/s00339-017-1319-1}

\begin{document}

\maketitle

\footnotetext[2]{ICube Laboratory, University of Strasbourg, CNRS UMR 7357, 300 Bd. Sébastien Brant, 67412 Illkirch, France}
\footnotetext[3]{ ECAM Strasbourg--Europe, 2 Rue de Madrid, 67012 Schiltigheim, France}
\footnotetext[1]{Corresponding author: robin.pierron@unistra.fr}

\begin{abstract}
We report on the first evidence of direct micro--peak machining using a photonic jet (PJ) with nanosecond laser pulses. PJ is a high concentrated propagative light beam with a full width at half maximum (FWHM) smaller than the diffraction limit. In our case, PJs are generated with a shaped optical fiber tip. Micro--peaks with a FWHM of around 1~$\mu$m, a height until 590~nm and an apex radius of 14~nm, were repeatability achieved on a silicon wafer. The experiments have been carried out in ambient air using a 100/140 multimode silica fiber with a shaped tip along with a 35~kHz pulsed laser emitting 100~ns pulses at 1064~nm. This study shows that the phenomenon occurs only at low energies, just under the ablation threshold. Bulk material appears to have moved around to achieve the peaks in a self--organized process. We hypothesize that the matter was melted and not vaporized; hydrodynamic flow of molten material governed by surface--tension forces may be the causes. This surface modification has many applications. For example, this paper reports on the decrease of wettability of a textured silicon wafer.   
\end{abstract}

\section{Introduction}
\label{intro}

Direct laser subwavelenght micromachining is currently a challeging topic. Photonic jets have already demonstrated the ability to reduce the laser etching size beyond the diffraction limit using micro--beads \cite{Abdurrochman,Munzer,Wu,Guo,Grojo,Mcleod} or more recently using shaped optical fiber tips \cite{Zelgowski,Pierron,Pierron2}. 
A Photonic jet (PJ) is a high concentrated propagative light beam with a full width at half maximum (FWHM) smaller than the diffraction limit \cite{Chen,Lecler,Itagi,Li}. The power density can be more than 200 times higher than the one of the incident wave \cite{Abdurrochman,Lecler,Heifetz}.
To achieve PJ, shaped fiber tips are obviously easier to move than a microsphere and therefore, to implement in an industrial process. Moreover, the fiber tips have no contact with the processed surface and are not altered by the removed material \cite{Zelgowski,Pierron,Pierron2}. Until now, only sub--micro PJ etching was reported. Micro--peaks formation was reported for processes using femtosecond laser pulses on thin metal and silicon films \cite{Her,Ivanov,Kuznetsov,Unger,Zhu,Kuznetsov2}. Micro--peaks were also generated using nanosecond laser pulses on thin gold film \cite{Moening,Moening2}, but never on silicon bulk.

In this paper, we report for the first time the possibility to achieve direct micro--peaks surface texturing using nanosecond pulses. Taking advantage of the photonic jet at the exist of a shaped optical fiber tip, peaks with a FWHM of around 1~$\mu$m, a height of almost an half micrometer and an apex radius of fews ten nanometers, were repeatability achieved on a silicon wafer. Surfaces with micro-peaks can have a wide range of applications. For example, A few millimeter squares surface with micro--spikes has shown an decrease of wettability.

\section{Experiment details}
\label{sec:1}

The laser source is a commercial near--infrared pulsed laser (VGEN ISP 1-40-30) emitting at 1064~nm with pulses of 100~ns at a repetition rate of 35~kHz. The output beam has a diameter of 6~mm at 1/e$^2$ and a beam quality factor (M$^2$) of 1.3. This corresponds to a quasi--Gaussian beam profile. An achromatic doublet, with a focal length of 19~mm, couples the laser beam into the fiber, whose position is controlled by XYZ microstages. 

\begin{figure*}
  	\includegraphics[width=1\textwidth]{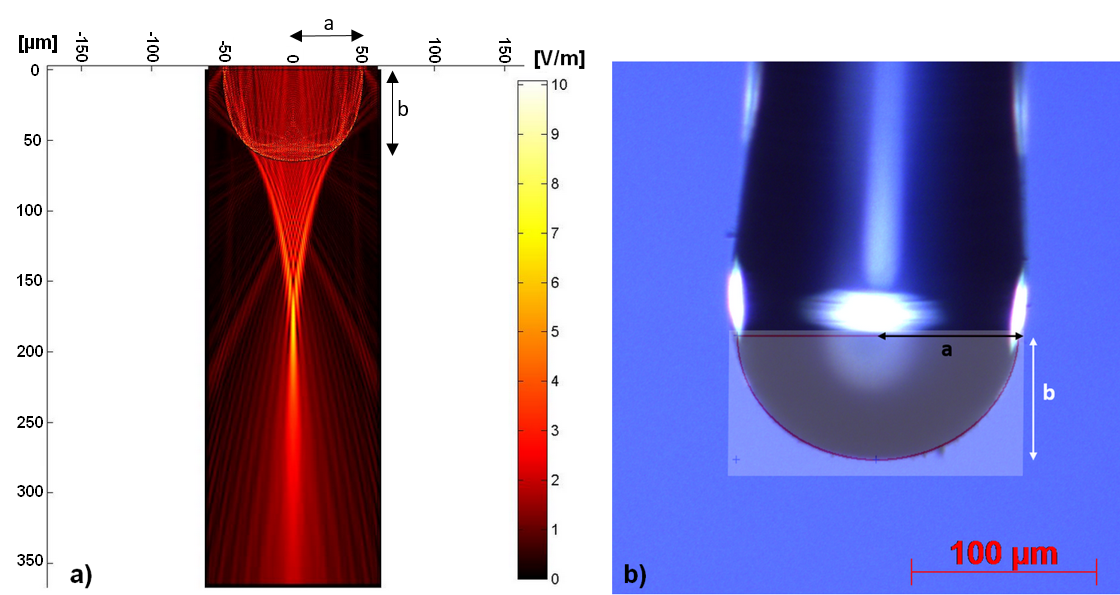}
	\caption{a) Simulation of the electrical field norm through and outside of the shaped silica fiber tip (a = 50 $\mu$m, b = 63 $\mu$m, $w_0$ = 1) excited by the fundamental mode of the fiber. b) Optical microscope view of the shaped fiber tip (a = 80 $\mu$m, b = 63 $\mu$m, $w_0$ = 1).}
	\label{fig:fiber_tip}       
\end{figure*}

The fiber system is a multimode step--index silica fiber with a core diameter of 100 $\mu$m and a cladding diameter of 140 $\mu$m. The numerical aperture (NA) is 0.22. The tip shape is numericaly described by a B\'ezier curve set by a base radius ($a$~=~50~$\mu$m), a tip length ($b$ = 63 $\mu$m) and a B\'ezier weight ($w_0$~=~1) \cite{Zelgowski}. It has been designed to achieve photonic jet of around 1 $\mu$m at a distance of around 100 $\mu$m of the tip when excited by the fundamenal mode (Fig.~\ref{fig:fiber_tip}~a). The numerical method is described in \cite{Zelgowski,Pierron,Pierron2}. The tip has been achieved by LovaLite using an electric--arc thermoforming technique \cite{Borsuk}. The result has a base radius $a$ slightly larger due the cladding: $a~\simeq$~80~$\mu m$ (Fig.~\ref{fig:fiber_tip}~b).

The shaped optical fiber tip is placed orthogonal to the sample. The fiber position is controlled by a Z--motorized stage and the sample position by XY--motorized stages. The distance between the tip and the sample is controlled by a camera with a 5x telecentric objective thanks to an image processing based on the reflectance of the tip into the sample. The laser, the motorized stages and the camera processing are controlled by LabView. The experiences are carried out at ambient atmosphere conditions. The power at the outputs of the laser and the fiber tip has been measured with a calorimeter Ophir Vega with a 12A--P sensor.  

In the following we name working distance the distance between the tip end and the maximum of intensity of the photonic jet. Experimentally, when both the laser pulse energy and the distance between the tip and the sample vary, the working distance corresponds to the distance for which the smallest mark is achieved with the smallest energy. It is intrinsic to the fiber tip and has been determined experimentally in our previous work to be 100 $\pm$ 2 $\mu$m for the tip shown in Fig. \ref{fig:fiber_tip}~b) \cite{Pierron2} not far from the 113 $\mu$m predicted by simulation (Fig.\ref{fig:fiber_tip}~a).
Our experiments have shown that this working distance is enough to avoid the re--deposition of melt material on the fiber tip if material is ablated. This also allows ensure the tip integrity during the motion control: no shock and easier control due the 2 $\mu$m PJ positioning tolerance. From simulations, we can give some general rules: (1) Photonic jet with the same FWHM (Full Width at Half Maximum) can be obtained at larger working distance using larger optical fiber core. However, in this case more energy is required to achieve the same process. Namely, it is not so easy to couple energy in the fundamental mode. (2) For a given fiber radius, generally the working distance is larger increasing the B\'ezier weight ($w_0$) or decreasing the tip length ($b$). However, the photonic jet FWHM increases. Therefore this tip was a good compromise. 

The sample is a monocrystalline silicon wafer with a passive layer (thickness of approximately 2 nm) and an initial roughness of 2 $\pm$ 1 nm. For each laser irradiation, 35 pulses have been used. Before and after the laser process, the sample was cleaned and dried with alcohol and dry air. 
 
The micro--peaks have been characterized with two different methods: a white light interferometric microscope and an atomic force microscope (AFM). The interferometric system based on coherence scanning interferometry was a Zygo NewView 7200 profilometer with an axial resolution of 3 nm and a lateral resolution of 550 nm (50x Mirau objective, NA~=~0.55). 
The AFM, a Park Systems XE--70 isolated inside an acoustic enclosure, worked in the non--contact mode. The field size used was 20x20 $\mu$m with a lateral resolution of 39 nm and a axial resolution of 2 nm.

The wettability of the flat and fakir's bed of nails--like textured sample was measured with a Kr\"uss Drop Shape Analyser DSA25. Static contact angle were measured with around 10 $\mu$L distilled water droplets by the sessile drop method. A computation method based on the analysis of the droplet shape determined by the model of Laplace--Young was used. 

\section{Results and discussion}
\label{sec:results_discussion}

\begin{figure*}
  	\includegraphics[width=1\textwidth]{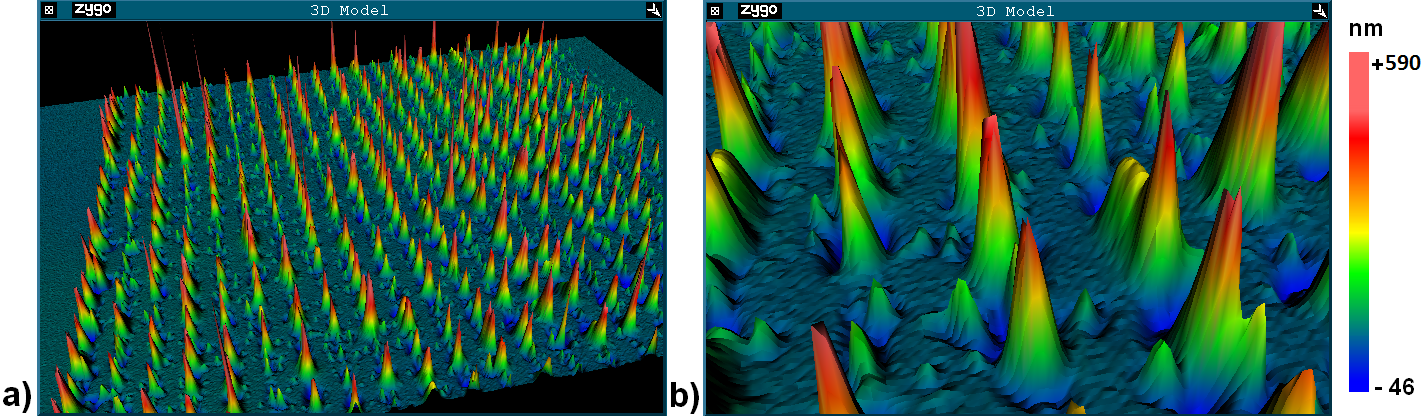}
	\caption{3D view with Zygo profilometer. a) 110x110 $\mu$m matrix on silicon with peaks every 5 $\mu$m; 35 pulses for each PJ peaks; pulse energy of 30~$\mu$J --- b) Zoom. Note that the unit lengths are not the same in the transverse plane and for height.}
	\label{fig:peaks_matrix}       
\end{figure*}

\begin{figure*}
  	\includegraphics[width=1 \textwidth]{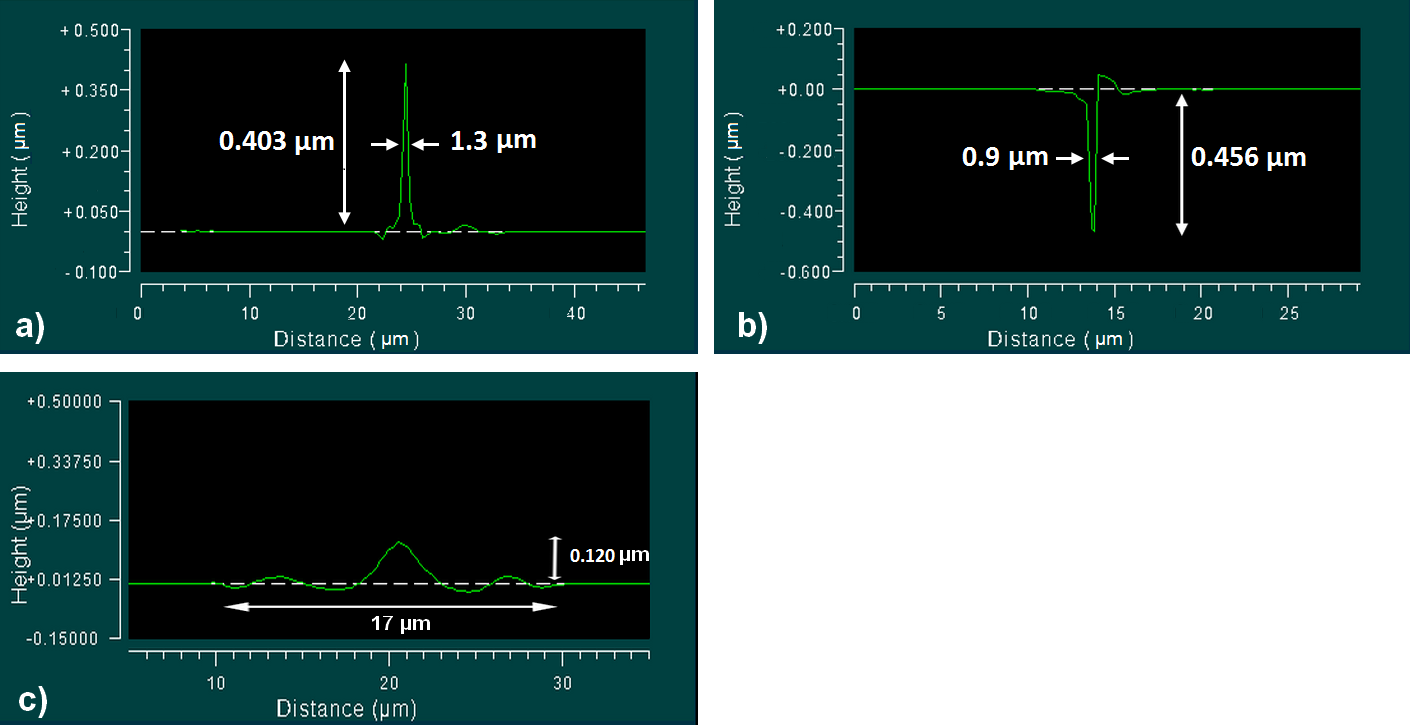}
	\caption{Examples of profiles on silicon with the Zygo microscope: a) a micro--peak; 35 pulses; pulse energy of 30 $\mu$J; tip-sample distance of 100 $\mu$m --- b) an ablation; 35 pulses; pulse energy of 36 $\mu$J; tip-sample distance of+ 100 $\mu$m --- c) damage area without peak; pulse energy of 30 $\mu$J; tip-sample distance of 110 $\mu$m. Note that the unit lengths are not the same in the x and y axes.}
	\label{fig:profile_zygo}       
\end{figure*}

A 110x100 $\mu$m matrix with peaks every 5 $\mu$m has been achieved on silicon. In Fig. \ref{fig:peaks_matrix}, the 3D view obtained by the interferometric microscope shows that micro--peak machining by PJ with an shaped optical fiber tip is a repeatable process. The measured mean height for 30 peaks was 354 $\pm$ 3 nm, the mean FWHM was 1 $\pm$ 0.6 $\mu$m and the maximum height was 590 $\pm$ 3 nm. Thus, peaks have a width larger than their height. Pay attention, due to the scales, the aspect ratio of the micro--peaks are different from what appears in Fig. \ref{fig:profile_zygo}. 

The PJ micro--peaks were obtained with 35 pulses, that corresponded to our mimium controlale number of pulses. The laser source has a minium repetion rate of 35 kHz and its minium controllabled shot time is 1~ms. The energy per pulse was 30~$\mu$J. Peak formation occurs only when the pulse energy is just under the ablation threshold (36 $\mu$J with our experimental conditions \cite{Pierron2}). For comparison purposes, the dimensions of a PJ ablation has been measured. An example of ablation with 35 pulses with a pulse energy of 36~$\mu$J has been achieved (cf. Fig. \ref{fig:profile_zygo}~b)). The PJ ablation is a sub--micro ablation with a deep of 456~$\pm$~3~nm and a FWHM of 0.9 $\pm$ 0.6 $\mu$m. An example of a PJ peak from the matrix is presented (cf. Fig. \ref{fig:profile_zygo}~a)). The PJ peak is a sub--micro peak with a height of 403~$\pm$~3~nm and a FWHM of 1.3~$\pm$~0.6~$\mu$m. Hence, the PJ peak has a FWHM with the same order magnitude (around 1~$\mu$m) as the FWHM of the PJ ablation, which is also the width of the PJ.

If a different distance between the tip and the sample is used, as illustrated in Fig. \ref{fig:profile_zygo}~c) with 110~$\mu$m, no more peaks are generated. The affect area (around 17~$\mu$m) has several maxima and minima and the highest maximum (around 120~nm) is not so high as the peaks and has a FWHM of 4 $\mu$m. This confirms the role of PJ in the peak generation.

\begin{figure*}
	\includegraphics[width=0.75\textwidth]{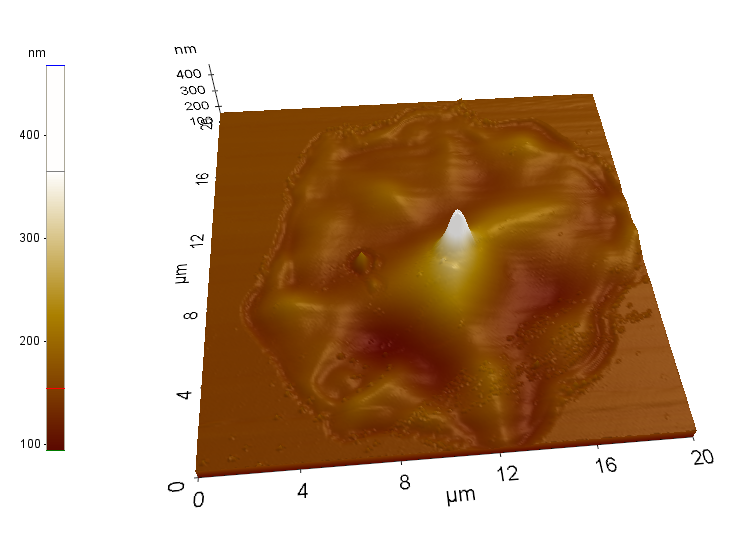}
  	\caption{3D view with the AFM. Example of a micro--peaks on silicon. Height of 335~$\pm$~2~nm, width 1.330~$\pm$~0.040~$\mu$m, apex radius of 14~$\pm$~2~nm -- 35 pulses -- pulse energy of 30 $\mu$J.}
  	\label{fig:peak_afm}       
\end{figure*}

\begin{figure}
  	\includegraphics[width=0.48\textwidth]{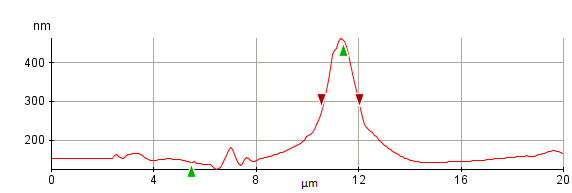}
	\caption{Example of a micro--peak profile on silicon with the AFM. Height of 335~$\pm$~2~nm, width of 1.330~$\pm$~0.040~$\mu$m, apex radius of 14~$\pm$~2~nm; 35 pulses; pulse energy of 30 $\mu$J.}
	\label{fig:peak_profile_afm}       
\end{figure}

In order to confirm the interferometer results, a micro--peak has been measured by AFM (cf. Fig. \ref{fig:peak_afm}). A height of 335~$\pm$~2~nm and a FWHM of 1.330 $\pm$ 0.040 $\mu$m have been measured (cf.~Fig.~\ref{fig:peak_profile_afm}). The micro--peak has a quasi-conical shape with an apex radius of 14~$\pm$~2~nm. Thus, with a small deviation (inferior to 10 \%), the AFM measurement confirms the interferometer profiles. 

The physical mechanisms of the micro--peak formation are hypothetical. A similar phenomenon is the micro--peak formation on thin metal films by femtosecond lasers \cite{Ivanov,Kuznetsov,Unger,Kuznetsov2}. The absorption processes are very different, however the properties of the PJ (characteristic sizes) allow interaction with a volume of matter similar to the case of femtosecond laser. In the two cases, the micro--peaks are generated by fluences just below the ablation threshold. Independently of the peaks, under nanosecond laser irradiation, melt dynamics is a dominant mechanism; femtosecond irradiation induces more complex dynamics including thermoplastic deformation \cite{Kuznetsov}. In silicon, as with femtosecond pulses, the formation mechanism of the micro--peaks could be apparently due to the hydrodynamic flow of molten material governed by surface--tension forces. The mechanism is due to the thermal expansion of the heated solid part of the material. This thermal extension induces stresses, which provide forces directly perpendicular to the solid--liquid interface. Theses forces push the melted material towards the center of the irradiated region. The peak formation would be arrested by the surface tensions forces and material solidification \cite{Kuznetsov,Kuznetsov2}.  

\begin{figure*}  	
  	\includegraphics[width=1\textwidth]{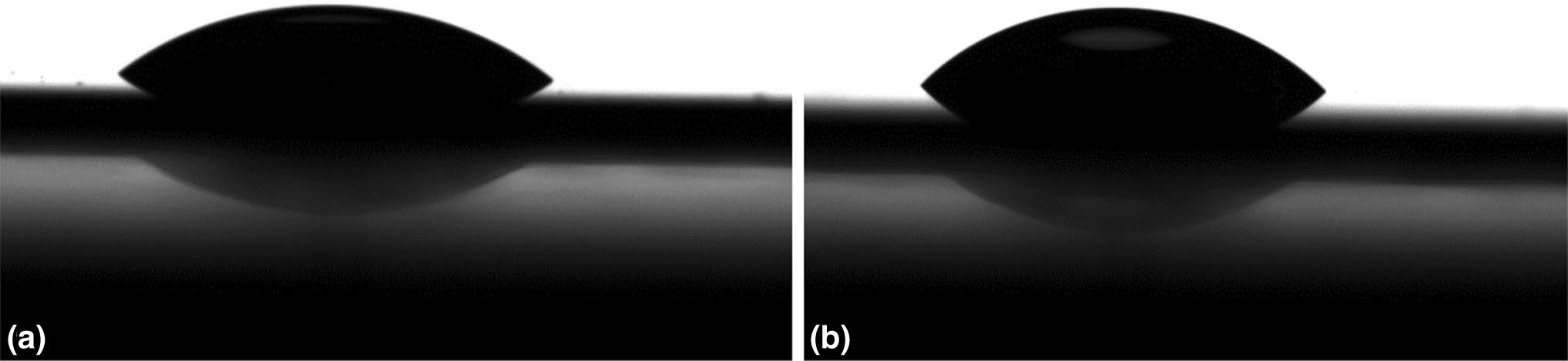}
	\caption{A drop of water ($\simeq$ 10 $\mu$L) on: a) a non--textured silicon wafer; contact angle of 39.3 $\pm$ 0.9 degrees -- b) the silicon wafer with micro-peaks, contact angle of 42.8 $\pm$ 0.8 degrees.}
	\label{fig:hydrophobia}       
\end{figure*}

An interest in microstructured surfaces with peaks has arisen thanks to the superhydrophobic effect \cite{Nayak,Hairaye}. As an application example, a 5 x 5 mm matrix with micro--peaks every 50 $\mu$m, called fakir's bed of nails surface, has been achieved on silicon. A  wettability test has been carried on the non--textured surface and textured surface (cf. Fig.~\ref{fig:hydrophobia}) with a drop of water. When the contact angle is below 90 degrees, the surface is considered hydrophilic. On the opposite, above 90 degrees the surface is hydrophobic \cite{Young,Wenzel,Cassie}. 
The contact angle of initial surface was 39.3 $\pm$ 0.9 degrees, whereas the one of the textured surface has increased to 42.8 $\pm$ 0.8 degrees. The silicon surface has been become less hydrophilic.

\section{Conclusion} 

Direct micro--spikes machining by photonic jet using nanosecond laser pulses have been observed. Repeatable micro--peaks have been fabricated by 35 pulses of 100 nanoseconds at 1064~nm with a photonic jet generated in the vicinity of an shaped optical fiber tip. They are obtained on silicon with 30~$\mu$J per pulses slightly under the ablation threshold (36~$\mu$J). An hypothesis is that micro--peaks are formed due to hydrodynamic flow of molten material governed by surface--tension forces. This will be investigated. The potential of these micro--peaks for reducing the hydrophilicity of silicon has been illustrated.

\paragraph{Acknowledgment}

The authors are grateful to Camille Hairaye (ICube Laboratory, France) for her technical assistance in the use of the characterization system of wettability.

\bibliography{mybibfile}


\bibliographystyle{ieeetr} 

\end{document}